\documentclass[a4paper]{article}

\usepackage{INTERSPEECH2022}  

\usepackage{comment}
\usepackage{tikz}
\usepackage{color}
\usepackage{subfigure}      
\usepackage{amsmath}
\usepackage{tipa} 

\title{ Speech intelligibility of simulated hearing loss sounds and\\ its prediction using the Gammachirp Envelope Similarity Index (GESI)}

\name{ Toshio Irino ,  Honoka Tamaru , and Ayako Yamamoto  }
\address{Faculty of Systems Engineering, Wakayama University, Japan
  }
\email{irino@wakayama-u.ac.jp, \{honoka.tamaru,yamamoto.ayako\}@g.wakayama-u.jp}
\begin{document}
\maketitle
\begin{abstract}
\vspace{-3pt}

In the present study, speech intelligibility (SI) experiments were performed using simulated hearing loss (HL) sounds in laboratory and remote environments to clarify the effects of peripheral dysfunction. Noisy speech sounds were processed to simulate the average HL of 70- and 80-year-olds using Wadai Hearing Impairment Simulator (WHIS). These sounds were presented to normal hearing (NH) listeners whose cognitive function could be assumed to be normal. The results showed that the divergence was larger in the remote experiments than in the laboratory ones. However, the remote results could be equalized to the laboratory ones, mostly through data screening using the results of tone pip tests prepared on the experimental web page. In addition, a newly proposed objective intelligibility measure (OIM) called the Gammachirp Envelope Similarity Index (GESI) explained the psychometric functions in the laboratory and remote experiments fairly well. GESI has the potential to explain the SI of HI listeners by properly setting HL parameters.

\end{abstract}


\vspace{-5pt}
\section{Introduction}
\vspace{-3pt}
As many countries approach super-aging society status, it is becoming crucial to develop next-generation assistive devices that compensate for individual listeners’ hearing difficulties. 
For this purpose, the hearing characteristics of elderly hearing impaired (HI) listeners should be measured and clarified effectively.
Although many psychometric experiments have been developed using relatively simple stimulus sounds, such as sinusoids and noise \cite{moore2013introduction}, several problems remain, especially in speech perception.
First, subjective listening tests are time consuming and costly because they are performed individually in a laboratory soundproof room with well-controlled equipment. Moreover, the cognitive load of long experiments is heavy for  HI listeners.  As a result, it is not easy to collect various data from a massive number of participants. Furthermore, the recent novel coronavirus outbreak has made it more difficult to conduct such formal experiments. More fundamentally, there is huge variability between elderly HI listeners in terms of audiograms as well as cognitive factors. It is not easy to distinguish the declined factors of the periphery, the auditory pathway, and cognition. 

The aim of this study is to answer the problem how to collect sufficiently large data to clarify and model the effect of the peripheral dysfunction separately from the pathway and cognitive factors. One approach is to perform experiments using a hearing loss (HL) simulator for normal hearing (NH) listeners whose cognitive function could be assumed to be normal.\cite{baer1993effects, stone1999tolerable, irino2013accurate, matsui2016effect}. We have developed a Wadai (or Wakayama University) Hearing Impairment Simulator, WHIS \cite{irino2013accurate, nagae2014hearing, irino2021whis,irino2022whis} which synthesizes reasonably high quality sounds that are applicable to speech perception experiments but not restricted to speech intelligibility (SI) \cite{irino2020speech}.
Another approach is to perform remote experiments using a crowdsourcing service, which enabled the collection of data with low cost and in a relatively short time, although the reliability should be carefully verified \cite{cooke2011crowdsourcing, 
paglialonga2020automated,padilla2021binaural,yamamoto2021comparison}.

It was also essential to develop an effective objective intelligibility measure (OIM) that could predict the SI of HI listeners whose hearing levels were individually different. Many OIMs have been proposed to evaluate speech enhancement and noise reduction algorithms for improving SI \cite{falk2015objective, van2018evaluation, yamamoto2020gedi};
however, most of them, except for a few examples such as HASPI \cite{kates2005coherence}, do not deal with the HL evaluation.
The experimental results for the SI of simulated hearing-loss sounds served as good initial tests for a new OIM.


In the present study, we performed SI experiments using WHIS in both laboratory and remote environments to clarify the effects of peripheral dysfunction. We found an effective data-screening method for remote experiments. We also proposed a new OIM to explain the experimental results.
	
\color{black}
\vspace{-5pt}
\section{Experiments}
\label{sec:Experiment}
%

\subsection{Speech sound source}
\label{sec:SourceSpeech}
\vspace{-3pt}

The speech sounds used for the subjective listening experiments were Japanese 4-mora words. They were uttered by a male speaker (label ID: mis) and drawn from a database of familiarity-controlled word lists, FW07 \cite{Kondo2007NTTTohoku}, which had been used in the previous experiments \cite{yamamoto2020gedi, yamamoto2021comparison}.
The dataset contained 400 words per each of the four familiarity ranks, and the average duration of a 4-mora word was approximately 700\,ms. The source speech sounds were obtained from the word set with the least familiarity to prevent increment of the SI score by guessing commonly used words.
Babble noise was added to the clean speech to obtain noisy speech sounds, which were referred to as unprocessed. The SNR conditions ranged from $-3$\,dB to $+9$\,dB in $3$-dB steps.
All sounds were processed in a sampling rate at 48\,kHz.

\vspace{-5pt}
\subsection{Hearing loss simulation}
\label{sec:HearingLossSimulation}
\vspace{-3pt}
Unprocessed noisy sounds were degraded to simulate HL using a new version of WHIS (${\rm WHIS_{v30}}$) \cite{irino2022whis}. Briefly, WHIS initially analyzes input sounds with the latest version of the gammachirp auditory filterbank (${\rm GCFB_{v23}}$), which calculates excitation patterns (EPs) based on the audiogram and the compression health, which represents the degree of  health in the compressive input-output (IO) function, of a HI listener. Then, WHIS synthesizes simulated hearing-loss sounds from the EPs using either of a direct time-varying filter  (DTVF) or a filterbank analysis/synthesis  (FBAS) method. The output sound distortion was smaller with WHIS than with the Cambridge version of the HL simulator \cite{baer1993effects, stone1999tolerable, nejime1997simulation}, which is used in Clarity-Prediction-Challenge \cite{claritychallenge}. 

The DTVF method was used in this study because the sound quality in preliminary listening was slightly better with it than with the FBAS method.
Table\,\ref{tab:HL} shows the average hearing levels used for the simulation: 70-year-old male listeners (hereafter 70yr), as defined in \cite{iso7029}, and 80-year-old listeners (hereafter 80yr), as defined in \cite{tsuiki2002nihon}.
The compression health is closely related to loudness recruitment \cite{moore2013introduction}. We set it to 0.5 to simulate moderate dysfunction in the IO function.
Moreover, we added a condition in which the sound pressure level (SPL) of the source sounds were simply reduced for 20\,dB (hereafter -20dB) to clarify the effects of SPL and high-frequency HL on SI. This reduction level was selected because the simulated 80yr sounds were approximately 20\,dB smaller than the source sounds. 


\begin{table}[t]
 \caption{Average hearing levels of 70 year-old male listeners \cite{iso7029} and 80 year-old listeners \cite{tsuiki2002nihon}.}\label{tab:HL}
   \vspace{-10pt}
 \centering
  \begin{tabular}{|c||c|c|c|c|c|c|c|c|}
   \hline
    Freq.& 125 & 250 & 500 & 1000 & 2000 & 4000 & 8000\\
   \hline \hline 
    70yr & 8 & 8 & 9 & 10 & 19  & 43 & 59 \\
    \hline 
     80yr & 24 & 24 & 27 & 28 & 33 & 48 & 69 \\
    \hline
  \end{tabular}
  \vspace{-10pt}
\end{table}

\vspace{-5pt}
\subsection{Experimental procedure}
\label{sec:ExpProcedure}
\vspace{-3pt}
We had developed a set of web pages that were usable in both laboratory and remote SI experiments\cite{yamamoto2021comparison}. Google Chrome was designated as a usable browser because it properly plays  48-kHz and 16-bit wav files on Windows and Mac systems. The participants were required to read information about the experiments before giving informed consent by clicking the agreement button twice in order to be transferred to the questionnaire page, which contained questions about age, type of wired-headphones or wired-earphones (the use of Bluetooth or a loudspeaker was not permitted), and native language (Japanese or not) as well as self-report of HL (yes or no). Then, they took tone pip tests, as described in Sec. \ref{sec:TonePip} and  \cite{yamamoto2022intelligibility}.
Next, the participants completed a training session in which they performed a very easy task using the same procedure as in the test sessions to familiarize themselves with the experimental tasks. The speech sounds were drawn from words with the highest familiarity rank and with an SNR above 0\,dB.

The main experimental pages were essentially the same as the previous ones \cite{yamamoto2021comparison}. The participants were instructed to write down the words they heard using hiragana during four-second periods of silence between words. 
The total number of presented stimuli was 400 words, comprising a combination of four HL conditions \{unprocessed, 70yr, 80yr, and -20dB \} and five SNR conditions with 20 words per condition. Each subject listened to a different word set, which was  randomly assigned to avoid bias caused by word difficulty. 
The experiment was divided into two one-hour tasks to fulfill the crowdsourcing requirement of the task duration.

We introduced several good practices to improve the quality of the experiments including measuring the listeners' vocabulary size. The others are described below. 

\vspace{-5pt}
\subsubsection{Leading sentence for familiarizing sound level}
\label{sec:LeadingSentence}
\vspace{-3pt}
We introduced the following leading sentence in each session: ``Speech sounds will be presented at this volume,'' in Japanese. This was followed by 10 test words.  The sentence and the words were processed using the same HL condition. In the preliminary experiments, when the words were randomly presented with various sound levels, it was not easy for listeners to concentrate on the sounds because they tried to avoid being frightened by a louder sound just after a softer sound. As a result, the SI scores in the unprocessed condition were lower than those in the previous experiment \cite{yamamoto2021comparison}. Overall, the leading sentence helped the listeners concentrate.

\vspace{-5pt}
\subsubsection{Tone pip test for estimating listening conditions}
\label{sec:TonePip}
\vspace{-3pt}

We also introduced a web page to estimate how much the sound level was presented above the threshold, which was determined by listener's absolute threshold, ambient noise level, and audio device. A sequence of 15 tone pips with -5\,dB decreasing steps was presented to the listeners, who were asked to report the number of audible tone pips, $N_{pip}$. The tone frequencies were 500\,Hz, 1000\,Hz, 2000\,Hz, and 4000\,Hz to cover the speech range. See \cite{yamamoto2022intelligibility} for more details.  

\vspace{-5pt}
\subsection{Laboratory experiments}
\label{sec:ExpLab}
\vspace{-3pt}
In the laboratory experiments, the sounds were presented diotically via a DA converter (SONY, NW-A55) over headphones (SONY, MDR-1AM2).
The sound pressure level (SPL) of the unprocessed sounds was 65\,dB in ${L_{eq}}$, which was calibrated with an artificial ear (Br\"{u}el \& Kj\ae r, Type\,4153) and a sound level meter (Br\"{u}el \& Kj\ae r, Type\,2250-L).
Listeners were seated in a sound-attenuated room with a background noise level of approximately 26dB in $L_{\rm Aeq}$.
Thirteen young Japanese NH listeners (six males and seven females, aged 19–24 years old) participated in the experiments. The subjects were all naive to our SI experiments and had a hearing level of less than 20\,dB between 125\,Hz and 8,000\,Hz.


\vspace{-5pt}
\subsection{Remote experiments}
\label{sec:ExpRemote}
\vspace{-3pt}

The experimental tasks were outsourced to a crowdsourcing service provided by Lancers Co. Ltd. in Japan \cite{Lancers} as in \cite{yamamoto2021comparison}.
Any crowdworker could participate in the experimental task on a first-come-first-served basis. The participants were asked to perform the experiments in a quiet place 
and set their device volume to an easily listenable level for the unprocessed condition and to a tolerably listenable level for the -20dB condition.
Although it was difficult to more strictly control the listening conditions, the tone pip tests described in Sec. \ref{sec:TonePip} helped to estimate them. 
In total, 28 participants completed the two experimental tasks.  There was a large variety of participants ages -- from 22 to 66 years old -- and there were three self-reported HI listeners. Their native language was Japanese. 

\begin{figure*}[t] 
\vspace{-10pt}
\begin{minipage} {0.49\hsize}
    \centering
    \includegraphics[width = 0.9\columnwidth]
    {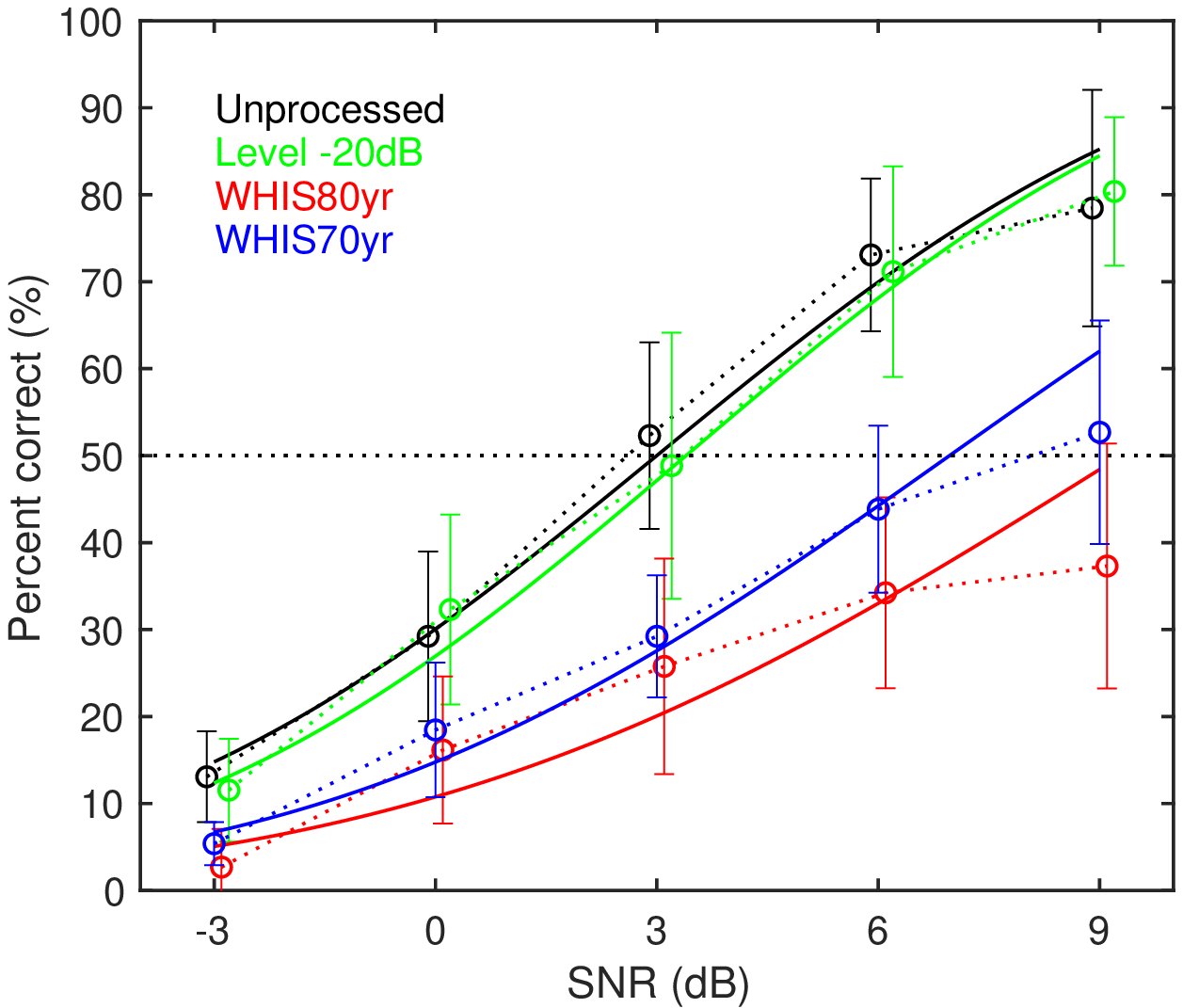}
\end{minipage}     
\begin{minipage} {0.49\hsize}    
    \includegraphics[width = 0.9\columnwidth]
    {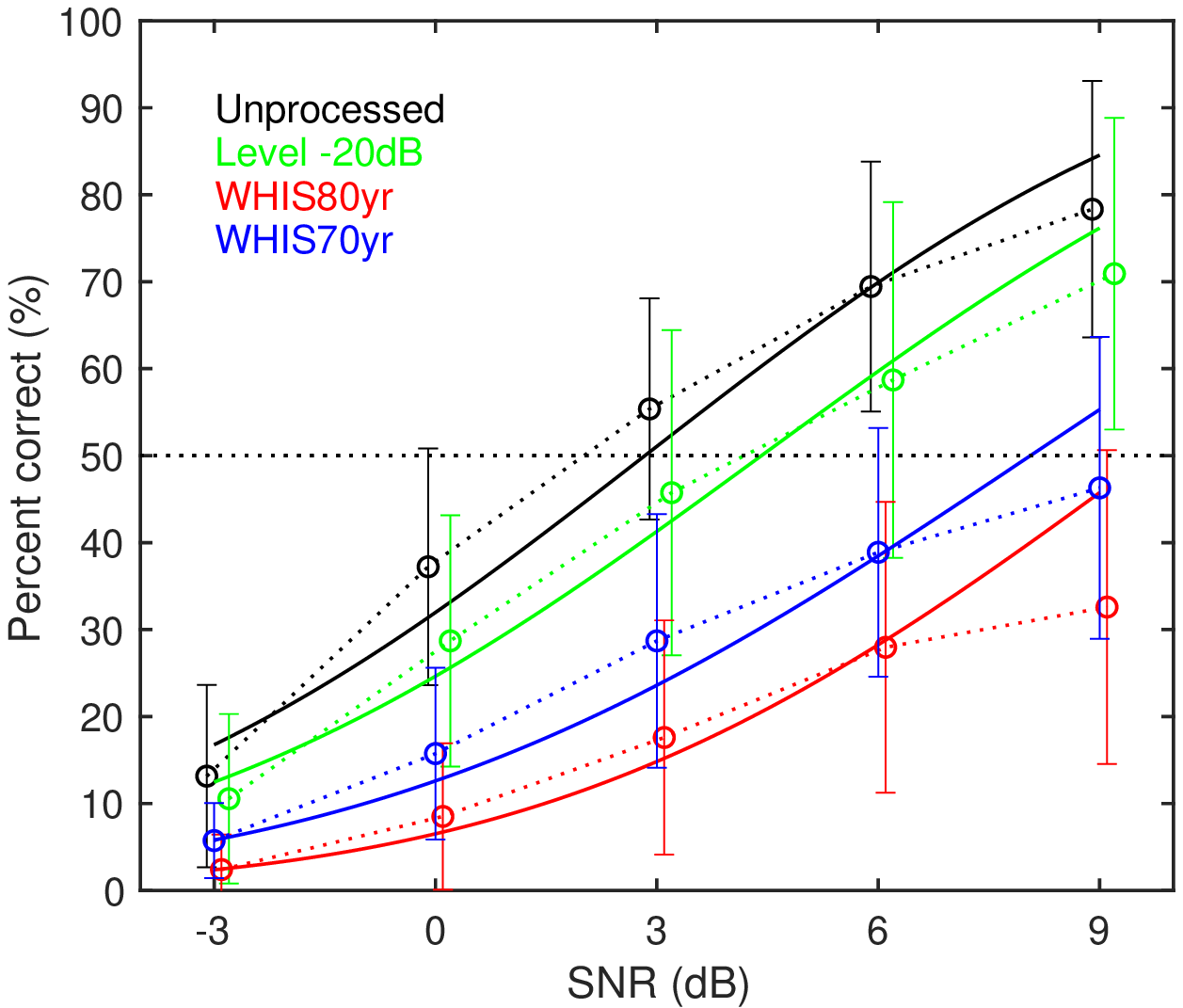} 
\end{minipage}  
\vspace{-5pt}
\caption{Mean and standard deviation (SD) of word correct rate (\%) across participants in the laboratory (left) and crowdsourced remote (right) experiments. Solid lines show the psychometric functions estimated with the cumulative Gaussian function. }
\label{fig:ExpLabRemote_PsychoFunc}
\vspace{-10pt}
\end{figure*}

\vspace{-5pt}
\subsection{Experimental results}
\label{sec:ExpResult}
\vspace{-3pt}
We performed data cleansing of participants' answers to calculate SI, as described in \cite{yamamoto2021comparison}, and we compared the results of the laboratory and remote experiment results.

\vspace{-5pt}
\subsubsection{Psychometric function of speech intelligibility}
\label{sec:SpIntel}
\vspace{-3pt}
Figure \ref{fig:ExpLabRemote_PsychoFunc} shows the word correct rates of the laboratory and remote experiments as a function of the SNR. Circles and error bars represent the mean and standard deviation (SD) across participants. The solid curves represent psychometric functions estimated with the cumulative Gaussian function using psignifit, a Bayesian method \cite{schutt2016painfree}. 

In the laboratory experiments (Fig. \ref{fig:ExpLabRemote_PsychoFunc}(left)), the psychometric functions of the unprocessed and -20dB conditions were almost the same. Therefore, the level reduction of 20\,dB did not affect SI in the well-controlled experiments with the young NH listeners. The psychometric functions of the 70yr and 80yr conditions were well below the unprocessed condition as expected.
The difference between the 80yr and -20dB conditions indicates that the HL in high frequencies seriously affected the SI, as a well-known fact. Listening difficulty corresponded to the severity of the simulated HL. The present study's results obtained here could represent the upper limits of the SI for elderly HI listeners whose hearing levels are similar to the 70yr and 80yr conditions.

The psychometric function of the unprocessed condition was almost the same in the remote experiments (right panel of Fig.\,\ref{fig:ExpLabRemote_PsychoFunc}) as in the laboratory experiments (left panel of Fig.\,\ref{fig:ExpLabRemote_PsychoFunc}). The order of the four lines from high to low was also the same. But the psychometric function of the -20dB condition was lower than that of the unprocessed condition, indicating that the  20\,dB gain reduction affected SI in the remote tests.  The SDs were also larger, as the psychometric functions of the individual listeners were very different. 
This was probably because the participants has different ages, hearing thresholds, and listening conditions, including ambient noise level.


\begin{figure}[t] 
\vspace{-10pt}
    \centering
    \hspace{-25pt}
    \includegraphics[width = 1.1\columnwidth]
    {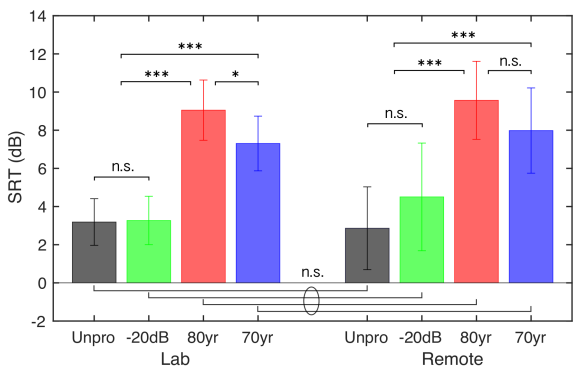} 
        \vspace{-13pt}
          \caption{Mean and SD of SRT (dB) across listeners for laboratory and remote tests.
          n.s.: not significant; ***: $p < 0.001$; *: $p < 0.05$.}
          \label{fig:SRT_LabRemote}
    \vspace{-15pt}
\end{figure}

\begin{figure}[t] 
    \centering
    \vspace{-10pt}
    \includegraphics[width = 0.8\columnwidth]
    {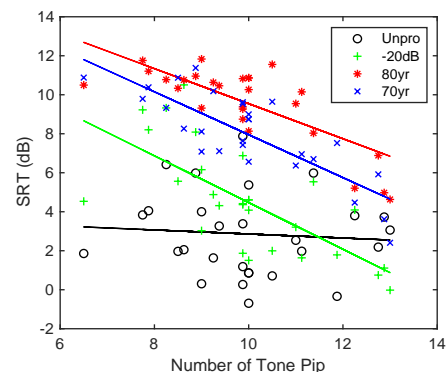} 
        \vspace{-8pt}
          \caption{Scatter plot for the mean reported number of audible tone pips, $N_{pip}$, vs. the mean SRT value (dB) in the remote experiments. Each point represents the individual listener. The solid lines show the regression results. }
          \label{fig:TonePipVsSRT_Remote}
    \vspace{-20pt}
\end{figure}

\vspace{-5pt}
\subsubsection{Speech reception threshold}
\label{sec:SRT}
\vspace{-3pt}
The speech reception threshold (SRT) was calculated as the SNR value at which the psychometric function reaches a 50 \% word correct rate. 
Figure \ref{fig:SRT_LabRemote} shows the mean and SD of the SRT values obtained in the laboratory and remote experiments for each HL condition.
There was a similar variance in the distribution of the mean values in these experiments.
 As a result of multiple comparison tests, there were no significant differences between the same HL conditions.
There was significant difference between the 70yr and 80yr conditions in the laboratory experiments but no significance in the remote experiments, probably because of the larger variability, as observed in the right panel of Fig \ref{fig:ExpLabRemote_PsychoFunc}.
The results of statistical significance tests were the same in the other combinations.

\vspace{-3pt}
\subsubsection{Data screening}
\label{sec:DataScreening}
\vspace{-3pt}
As described in Sec. \ref{sec:TonePip}, we introduced tone pip listening tests.  We surveyed the relationship between $N_{pip}$ and the SRT values for individual listeners.
Figure \ref{fig:TonePipVsSRT_Remote} shows a scatter plot between $N_{pip}$ averaged across four tone frequencies and the SRT value averaged across the four HL conditions in the remote experiments. 
There was no significant correlation in the unprocessed  ($ r=-0.079; p=0.70$) condition. However, there was a strongly significant correlation in the -20dB ($ r=-0.70; p=4.4\times10^{-5}$ ), 80 yr ($ r=-0.73; p=1.8\times10^{-5}$ ), and 70\, yr ($ r=-0.82; p=2.3\times10^{-7}$) conditions. This implies that the tone pip test may provide good information about the listening conditions, which would be useful for data screening. Please note that the average $N_{pip}$ was more than 11 in the laboratory experiments.

When $N_{pip}$ was less than 9, the dynamic range above the threshold was less than 40\,dB ($= 5\,{\rm dB\,step} \times (9-1)$). This was insufficient to recognize low-level consonants even in the -20dB condition which is a flat level reduction. The condition was obviously different from the well-controlled one in the laboratory. When we removed the participants whose $N_{pip}$ were less than 9, the mean SRTs for the 80yr conditions in Fig. \ref{fig:SRT_LabRemote} became significantly different from the 70yr condition ($p<0.05$). As such, the results were virtually the same between the remote and laboratory experiments. The remote experiments could be used as an alternative to strict laboratory experiments when data screening is performed using the tone pip tests.



\vspace{-5pt}
\section{Objective intelligibility measure (OIM)}
\label{sec:OIM}
The results described in Sec. \ref{sec:ExpResult} are a good initial test of OIMs to explain the SI of HI listeners.
Most of conventional OIMs \cite{falk2015objective, van2018evaluation}, except for a few, such as HASPI \cite{kates2005coherence}, normalize both of the reference and test sounds to the same rms level. 
Thus, theoretically, they cannot predict the difference between the
unprocessed and -20dB conditions in the remote experiments, as shown in the right panel of  Fig.\ref{fig:ExpLabRemote_PsychoFunc}. There was also inter-subject variability. The problems should be resolved in a desired OIM.
%
\vspace{-8pt}
\subsection{Gammachirp envelope similarity index (GESI)}
\vspace{-5pt}
We developed a new OIM called the Gammachirp Envelope Similarity Index (GESI,
[\textdyoghlig\'esi]),
based on a framework similar to the Gammachirp Envelope Distortion Index (GEDI, [\textdyoghlig\'eda\i]) \cite{yamamoto2020gedi}. The process started with the new version of GCFB, ${\rm GCFB_{v23}}$ with the NH setting in Sec. \ref{sec:HearingLossSimulation}  \cite{irino2022whis}. The next stages were envelope extraction and filtering with an IIR version of a modulation filterbank (MFB), initially introduced in sEPSM \cite{jorgensen2011predicting, jorgensen2013multi}. Then, we applied a new method to compare the MFB outputs between the reference ($m_{ij}^r(\tau)$) and the test ($m_{ij}^t(\tau)$). We used the following modified version of cosine similarity:
\setlength{\abovedisplayskip}{2pt} 
\setlength{\belowdisplayskip}{2pt} 
\begin{eqnarray}
  S_{ij} &=&
  \frac{\sum_{\tau} m_{ij}^r(\tau)\cdot 
  m_{ij}^t(\tau)} 
  {(\sum_{\tau}  {m_{ij}^r(\tau)}^2)^{\rho} \cdot 
    (\sum_{\tau} {m_{ij}^t(\tau)}^2)^{\,(1-\rho)}}
    \label{eq:similarity}
\end{eqnarray}
where $i$ is the GCFB channel, $j$ is the MFB channel, $\tau$ is a frame number, and $\rho$ $\{\rho \,|\, 0 \le \rho \le 1\} $ is a weight value that deals with the asymmetry in the levels of the reference and test sounds. When $\rho$ is 0.5, $S$ is the original cosine similarity. 
$\rho$ seemed to closely correlate with the dynamic range of speech sounds presented to the listeners.
As shown in Fig. \ref{fig:TonePipVsSRT_Remote}, the reported number of tone pips, $N_{pip}$, correlated with the mean SRT values and could be used to estimate the dynamic range. From the preliminary simulation, we assumed that $\rho$ was associated with $N_{pip}$ as follows:
\begin{eqnarray}
    \rho & = & 0.50 + 0.02\cdot(15 - N_{pip}).
      \label{eq:rho=05}
\end{eqnarray}
The similarity metric, $d_s$, is a weighted sum of $S_{ij}$ with weighting functions of  $w_i$ and $w_j$ as follows:
\begin{eqnarray}
   d_s & = & \frac{1}{MN} \sum_{i=1}^N \sum_{j=1}^M w_i\, w_j\, S_{ij}.
  \label{eq:metric}
  \vspace{-10pt}
\end{eqnarray}
%
%
It is important to properly choose weighting functions based on knowledge of speech perception. In the preliminary simulation, a uniform $w_i$ could not explain the results in Fig. \ref{fig:ExpLabRemote_PsychoFunc}.
This was probably because excitation patterns in the low frequencies were largely dominated by glottal pulse components, which were not important for phoneme identification, as indicated in the Articulation Index \cite{ANSI_S3-5_1969}.
We set $w_i$ as a Size-Shape Image (SSI) weight \cite{matsui2022modelling}, which was recently introduced to explain psychometric functions of size perception from speech sounds and was based on a computational theory of speech perception \cite{irino2002segregating}. We used the following equation:
\begin{eqnarray}
  w^{SSI}_i & = & \min(f_{p,i}/\bar F_o, h_{max})/h_{max}
  \label{eq:SSIweight}
\end{eqnarray}
where $f_{p,i}$ is the peak frequency of the $i$th GCFB channel and  $h_{max}$ is an upper limit parameter. In addition, $\bar Fo$ is a geometric mean of fundamental frequencies, $Fo$, of the reference sound, estimated by the WORLD speech synthesizer \cite{morise2016world}.  In contrast, $w_j$ is uniform in the current simulation but adjustable.

The metric value, $d_s$, was converted into word correct rate (\%) or intelligibility, $I$, by a sigmoid function used in STOI and ESTOI as $I = 100/(1+\exp(a\cdot d_s + b))$.

\vspace{-5pt}
\subsection{Simulation results}
\begin{figure}[t]
    \vspace{-5pt}
    \centering
    \hspace{-15pt}
    \begin{minipage}[b]{0.49\columnwidth}
    \includegraphics[width = 1.1\columnwidth]
    {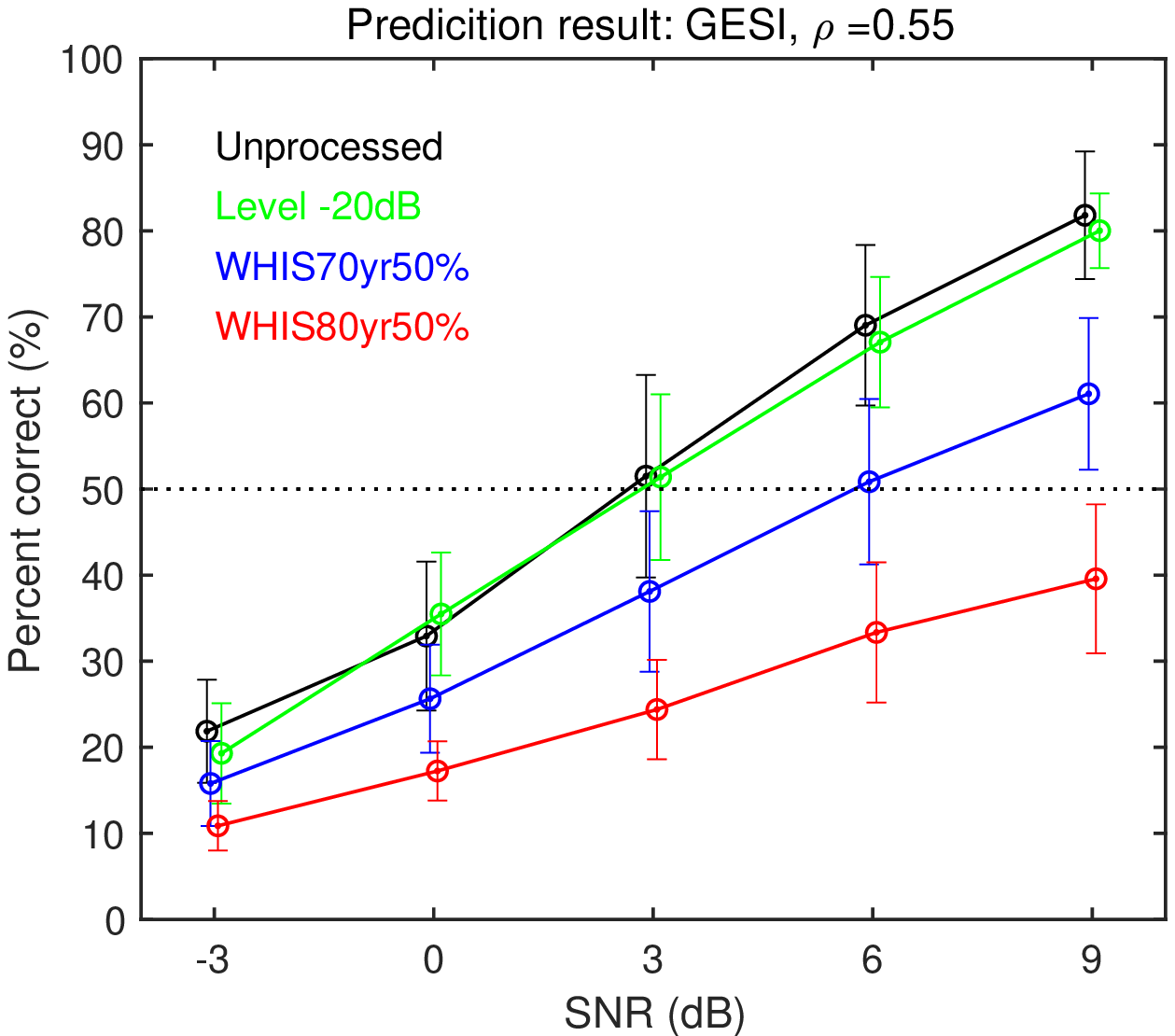} 
    \end{minipage}
    \begin{minipage}[b]{0.49\columnwidth}
    \includegraphics[width = 1.1\columnwidth]
    {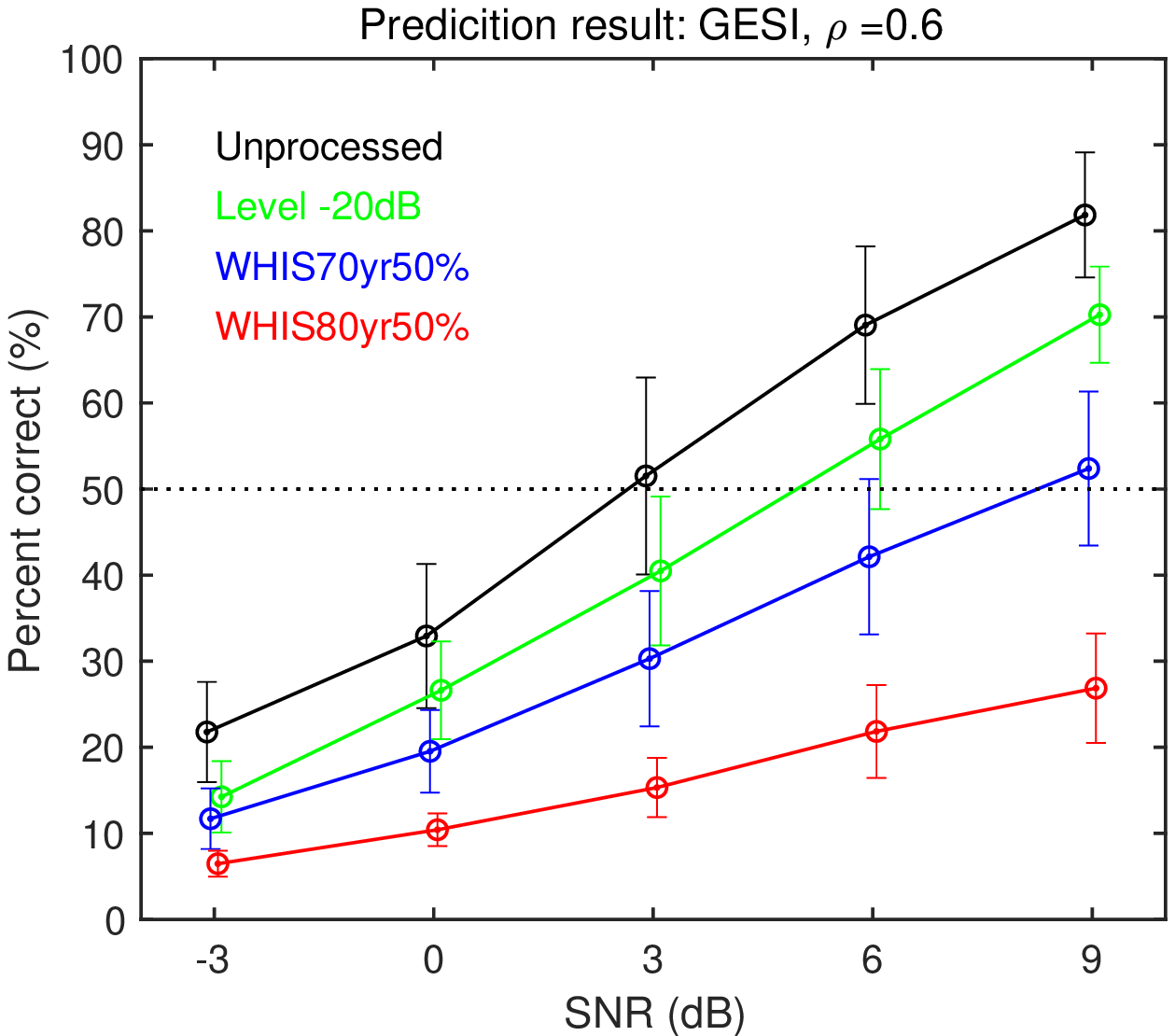} 
    \end{minipage}
        \vspace{-8pt}
          \caption{Prediction using GESI (left: $\rho=0.55$, right: $\rho=0.60$). The error bar represents the SD across words.   } \label{fig:GESI} 
    \vspace{-12pt}
\end{figure}

\vspace{-3pt}
The mean reported numbers of tone pips, $\bar N_{pip}$, were 12.5 for the laboratory experiments and 10.0 for the remote experiments (Fig.\ref{fig:TonePipVsSRT_Remote}). The coefficients, $\rho$, were calculated as 0.55 and 0.60 using Eq.\ref{eq:rho=05}.
Figure \ref{fig:GESI} shows the simulated psychometric functions using 20 words when the sigmoid function was determined to minimize the error in the unprocessed condition. The left and right panels are similar to those in Fig. \ref{fig:ExpLabRemote_PsychoFunc}. It was possible to control the location of the psychometric function of the -20dB condition by adjusting the parameter $\rho$. It was also possible to reasonably explain the 70yr and 80yr conditions. Since the hearing-loss simulation by WHIS demonstrated fairly good precision \cite{irino2022whis},  GESI has the potential to explain the SI of individual HI listeners by setting their audiograms and compression health to embedded GCFB.

\begin{figure}[t] 
    \centering
    \hspace{-10pt}
    \begin{minipage}[b]{0.49\columnwidth}
    \includegraphics[width = 1.1\columnwidth]
    {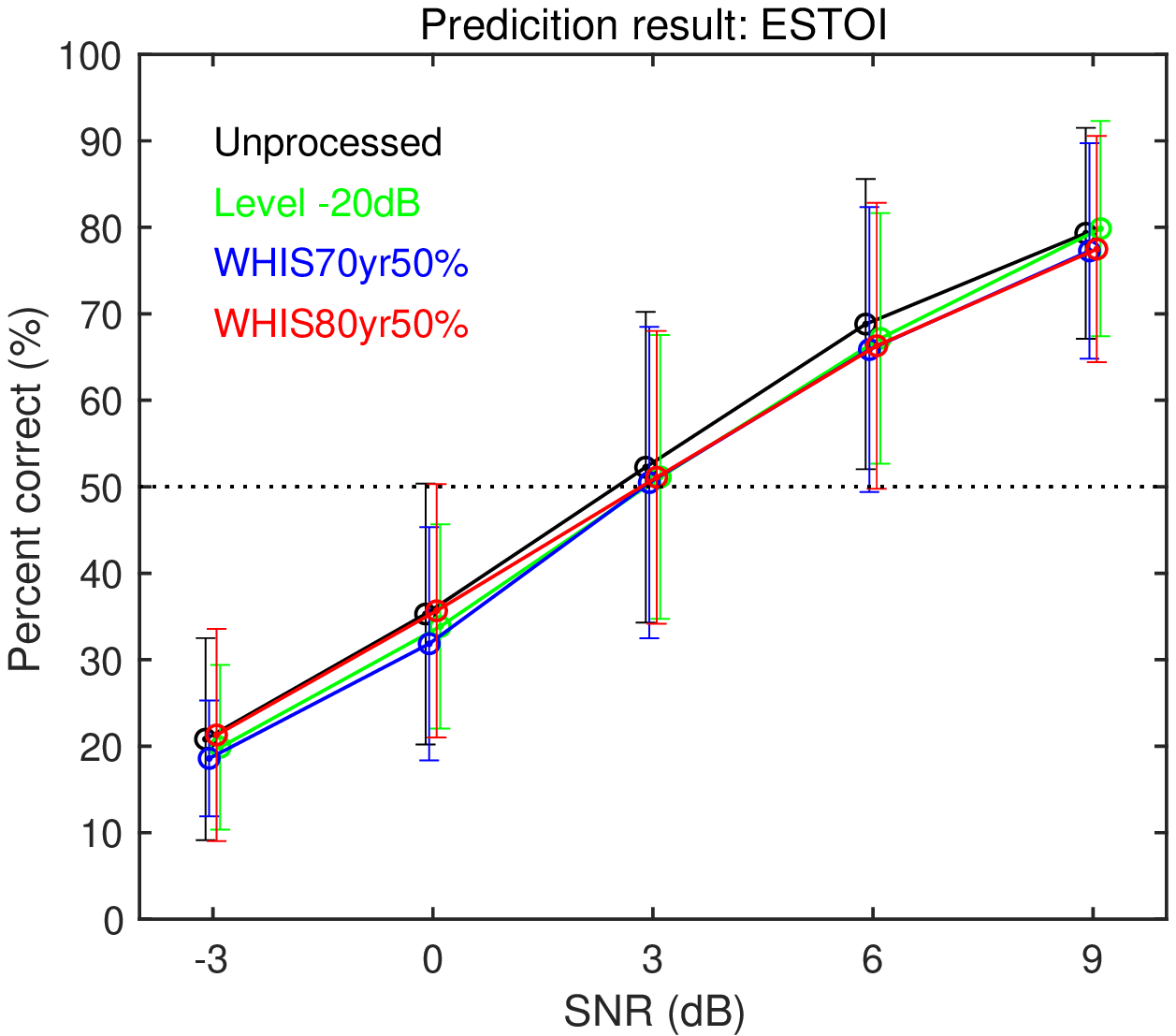} 
    \end{minipage}
    \begin{minipage}[b]{0.49\columnwidth}
    \includegraphics[width = 1.1\columnwidth]
    {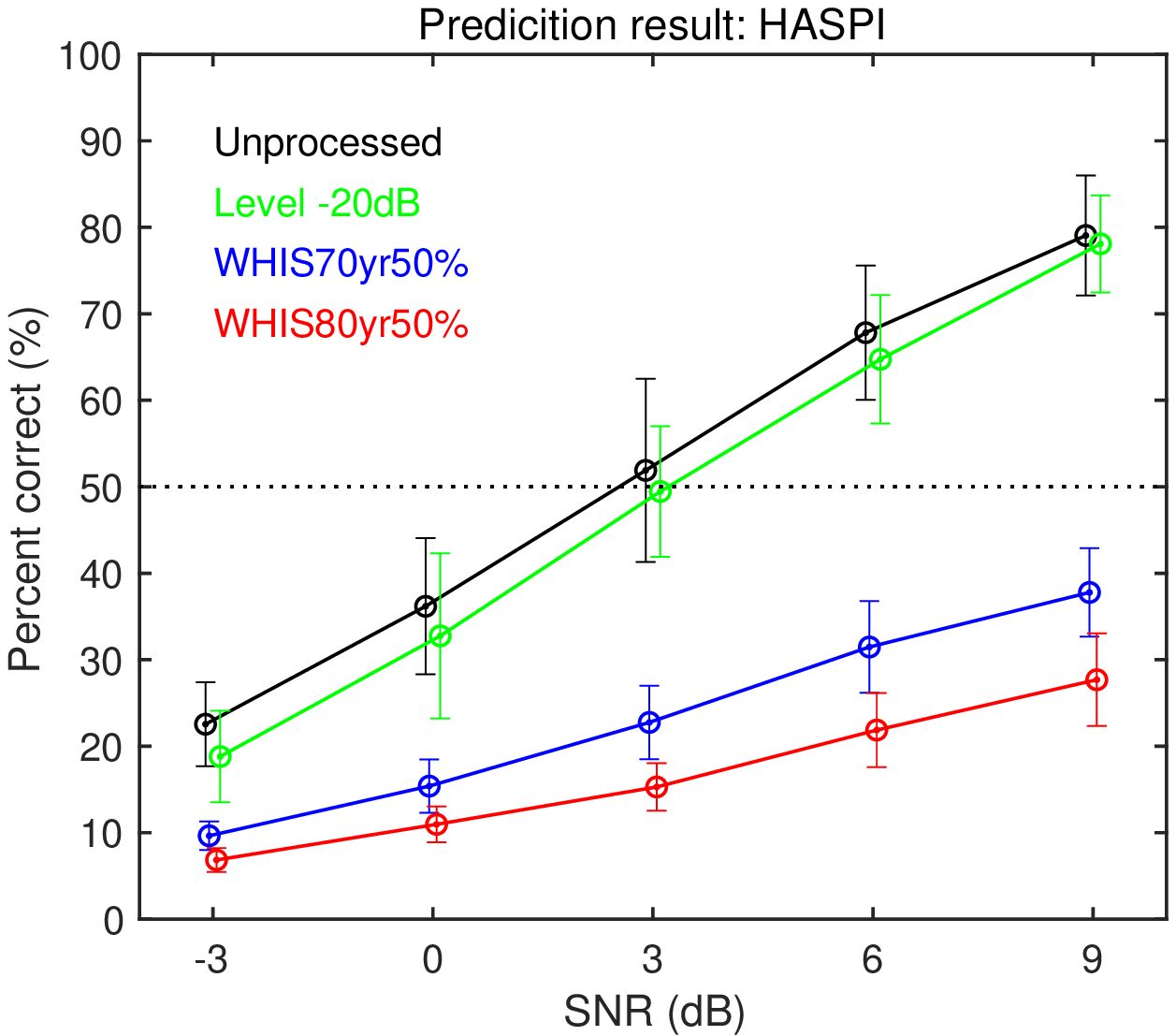} 
    \end{minipage}
        \vspace{-10pt}
          \caption{Prediction using ESTOI (left) and HASPI (right).  }
          \label{fig:ESTOI-HASPI}
    \vspace{-15pt}
\end{figure}

The left panel of Fig. \ref{fig:ESTOI-HASPI} shows the prediction results using ESTOI, which is one of the most popular OIMs. It was not possible to explain the subjective results as expected. The STOI results were almost as the ESTOI ones. The right panel of Fig. \ref{fig:ESTOI-HASPI} shows the prediction results using HASPI \cite{kates2005coherence}. Although the results were much better than those using ESTOI, the -20dB line was just below the unprocessed line. It was not possible to explain the results in the right panel of Fig.\,\ref{fig:ExpLabRemote_PsychoFunc} as well as the individual listeners' results because no method was proposed to control the location of the psychometric function.

\vspace{-5pt}
\section{Conclusions}
\vspace{-3pt}
In  the present  study,  we performed laboratory and remote experiments on SI in which NH listeners listened to simulated HL sounds produced by WHIS. The remote results could be equalized to the laboratory results, mostly through data screening using the self-reported number of audible tone pips. This method may enables us to utilize the remote experiments to collect a massive amount of SI data with less effort and time. Moreover, we developed GESI, a new OIM that was able to explain the current subjective results and has the potential to explain the SI of HI listeners. GESI is opened in our github repository\,\cite{github_amlab}.




\vspace{-5pt}
\section{Acknowledgements}
\vspace{-3pt}
This research was supported by JSPS KAKENHI Nos. 21H03468, and 21K19794.


  \newpage
  \eightpt
  \bibliographystyle{IEEEtran}
  \bibliography{Reference_InterSp22.bib}

\begin{thebibliography}{10}
\providecommand{\url}[1]{#1}
\csname url@samestyle\endcsname
\providecommand{\newblock}{\relax}
\providecommand{\bibinfo}[2]{#2}
\providecommand{\BIBentrySTDinterwordspacing}{\spaceskip=0pt\relax}
\providecommand{\BIBentryALTinterwordstretchfactor}{4}
\providecommand{\BIBentryALTinterwordspacing}{\spaceskip=\fontdimen2\font plus
\BIBentryALTinterwordstretchfactor\fontdimen3\font minus \fontdimen4\font\relax}
\providecommand{\BIBforeignlanguage}[2]{{%
\expandafter\ifx\csname l@#1\endcsname\relax
\typeout{** WARNING: IEEEtran.bst: No hyphenation pattern has been}%
\typeout{** loaded for the language `#1'. Using the pattern for}%
\typeout{** the default language instead.}%
\else
\language=\csname l@#1\endcsname
\fi
#2}}
\providecommand{\BIBdecl}{\relax}
\BIBdecl

\bibitem{moore2013introduction}
B.~C.~J. Moore, \emph{An introduction to the psychology of hearing}, 6th~ed.\hskip 1em plus 0.5em minus 0.4em\relax Leiden, The Netherlands: Brill, 2013.

\bibitem{baer1993effects}
T.~Baer and B.~C. Moore, ``Effects of spectral smearing on the intelligibility of sentences in noise,'' \emph{The Journal of the Acoustical Society of America}, vol.~94, no.~3, pp. 1229--1241, 1993.

\bibitem{stone1999tolerable}
M.~A. Stone and B.~C. Moore, ``Tolerable hearing aid delays. i. estimation of limits imposed by the auditory path alone using simulated hearing losses,'' \emph{Ear and Hearing}, vol.~20, no.~3, pp. 182--192, 1999.

\bibitem{irino2013accurate}
T.~Irino, T.~Fukawatase, M.~Sakaguchi, R.~Nisimura, H.~Kawahara, and R.~D. Patterson, ``Accurate estimation of compression in simultaneous masking enables the simulation of hearing impairment for normal-hearing listeners,'' in \emph{Basic Aspects of Hearing}.\hskip 1em plus 0.5em minus 0.4em\relax Springer, 2013, pp. 73--80.

\bibitem{matsui2016effect}
T.~Matsui, T.~Irino, M.~Nagae, H.~Kawahara, and R.~D. Patterson, ``The effect of peripheral compression on syllable perception measured with a hearing impairment simulator,'' in \emph{Physiology, Psychoacoustics and Cognition in Normal and Impaired Hearing}.\hskip 1em plus 0.5em minus 0.4em\relax Springer, Cham, 2016, pp. 307--314.

\bibitem{nagae2014hearing}
M.~Nagae, T.~Irino, R.~Nisimura, H.~Kawahara, and R.~D. Patterson, ``Hearing impairment simulator based on compressive gammachirp filter,'' in \emph{Signal and Information Processing Association Annual Summit and Conference (APSIPA), 2014 Asia-Pacific}.\hskip 1em plus 0.5em minus 0.4em\relax IEEE, 2014, pp. 1--4.

\bibitem{irino2021whis}
\BIBentryALTinterwordspacing
T.~Irino, ``A new implementation of hearing impairment simulator {WHIS} based on the gammachirp auditory filterbank,'' in \emph{The 3rd Japan-Taiwan Symposium on Psychological and Physiological Acoustics, in Japanese}, vol.~51, no.~8.\hskip 1em plus 0.5em minus 0.4em\relax Acoust. Soc. Japan, 2021, pp. 545--550. [Online]. Available: \url{https://github.com/AMLAB-Wakayama/WHIS}
\BIBentrySTDinterwordspacing

\bibitem{irino2022whis}
\BIBentryALTinterwordspacing
------, ``{WHIS}: Hearing impairment simulator based on the gammachirp auditory filterbank,'' \emph{arXiv preprint}, vol. arXiv:2206.06604, 2022. [Online]. Available: \url{https://doi.org/10.48550/arXiv.2206.06604}
\BIBentrySTDinterwordspacing

\bibitem{irino2020speech}
T.~Irino, S.~Higashiyama, and H.~Yoshigi, ``Speech clarity improvement by vocal self-training using a hearing impairment simulator and its correlation with an auditory modulation index.'' in \emph{Interspeech 2020}, 2020, pp. 2507--2511.

\bibitem{cooke2011crowdsourcing}
M.~Cooke, J.~Barker, M.~L.~G. Lecumberri, and K.~Wasilewski, ``Crowdsourcing for word recognition in noise,'' in \emph{Interspeech 2011, Twelfth Annual Conference of the International Speech Communication Association}, 2011.

\bibitem{paglialonga2020automated}
A.~Paglialonga, E.~M. Polo, M.~Zanet, G.~Rocco, T.~van Waterschoot, and R.~Barbieri, ``An automated speech-in-noise test for remote testing: Development and preliminary evaluation,'' \emph{American Journal of Audiology}, vol.~29, no.~3S, pp. 564--576, 2020.

\bibitem{padilla2021binaural}
\BIBentryALTinterwordspacing
A.~Padilla-Ortiz and F.~Ordu{\~n}a-Bustamante, ``Binaural speech intelligibility tests conducted remotely over the internet compared with tests under controlled laboratory conditions,'' \emph{Applied Acoustics}, vol. 172, p. 107574, 2021. [Online]. Available: \url{https://doi.org/10.1016/j.apacoust.2020.107574}
\BIBentrySTDinterwordspacing

\bibitem{yamamoto2021comparison}
\BIBentryALTinterwordspacing
A.~Yamamoto, T.~Irino, K.~Arai, S.~Araki, A.~Ogawa, K.~Kinoshita, and T.~Nakatani, ``Comparison of remote experiments using crowdsourcing and laboratory experiments on speech intelligibility,'' in \emph{Interspeech 2021}, 2021, pp. 181--185. [Online]. Available: \url{https://www.doi.org/10.21437/Interspeech.2021-174}
\BIBentrySTDinterwordspacing

\bibitem{falk2015objective}
\BIBentryALTinterwordspacing
T.~H. Falk, V.~Parsa, J.~F. Santos, K.~Arehart, O.~Hazrati, R.~Huber, J.~M. Kates, and S.~Scollie, ``Objective quality and intelligibility prediction for users of assistive listening devices: Advantages and limitations of existing tools,'' \emph{IEEE signal processing magazine}, vol.~32, no.~2, pp. 114--124, 2015. [Online]. Available: \url{https://ieeexplore.ieee.org/abstract/document/7038268/}
\BIBentrySTDinterwordspacing

\bibitem{van2018evaluation}
S.~Van~Kuyk, W.~B. Kleijn, and R.~C. Hendriks, ``An evaluation of intrusive instrumental intelligibility metrics,'' \emph{IEEE/ACM Transactions on Audio, Speech, and Language Processing}, vol.~26, no.~11, pp. 2153--2166, 2018.

\bibitem{yamamoto2020gedi}
\BIBentryALTinterwordspacing
K.~Yamamoto, T.~Irino, S.~Araki, K.~Kinoshita, and T.~Nakatani, ``{GEDI}: Gammachirp envelope distortion index for predicting intelligibility of enhanced speech,'' \emph{Speech Communication}, vol. 123, pp. 43--58, 2020. [Online]. Available: \url{https://doi.org/10.1016/j.specom.2020.06.001}
\BIBentrySTDinterwordspacing

\bibitem{kates2005coherence}
J.~M. Kates and K.~H. Arehart, ``{Coherence and the speech intelligibility index},'' \emph{The Journal of the Acoustical Society of America}, vol. 117, no. 4 Pt 1, pp. 2224--2237, 2005.

\bibitem{Kondo2007NTTTohoku}
\BIBentryALTinterwordspacing
K.~Kondo, S.~Amano, Y.~Suzuki, and S.~Sakamoto, ``{NTT}-{T}ohoku university familiarity-controlled word lists 2007 ({FW07}),'' 2007. [Online]. Available: \url{http://research.nii.ac.jp/src/en/FW07.html}
\BIBentrySTDinterwordspacing

\bibitem{nejime1997simulation}
Y.~Nejime and B.~C. Moore, ``Simulation of the effect of threshold elevation and loudness recruitment combined with reduced frequency selectivity on the intelligibility of speech in noise,'' \emph{The Journal of the Acoustical Society of America}, vol. 102, no.~1, pp. 603--615, 1997.

\bibitem{claritychallenge}
\BIBentryALTinterwordspacing
C.~challenge commitee, ``Clarity challenge - prediction challenge,'' 2022, (Last: 26 Feb. 2022). [Online]. Available: \url{http://claritychallenge.org}
\BIBentrySTDinterwordspacing

\bibitem{iso7029}
\BIBentryALTinterwordspacing
I.~7029:2017, ``Acoustics — statistical distribution of hearing thresholds related to age and gender,'' \emph{ISO}, 2017. [Online]. Available: \url{https://www.iso.org/standard/42916.html}
\BIBentrySTDinterwordspacing

\bibitem{tsuiki2002nihon}
T.~Tsuiki, S.~Sasamori, Y.~Minami, T.~Ichinohe, K.~Murai, S.~Murai, and H.~Kawashima, ``Age effect on hearing: a study on japanese,'' \emph{Audiology Japan (in Japanese)}, vol.~45, no.~3, pp. 241--250, 2002.

\bibitem{yamamoto2022intelligibility}
\BIBentryALTinterwordspacing
A.~Yamamoto, T.~Irino, S.~Araki, K.~Arai, A.~Ogawa, K.~Kinoshita, and T.~Nakatani, ``Subjective intelligibility of speech sounds enhanced by ideal ratio mask via crowdsourced remote experiments with effective data screening,'' \emph{arXiv Preprint}, vol. arXiv:2203.16760, 2022. [Online]. Available: \url{https://doi.org/10.48550/arXiv.2203.16760}
\BIBentrySTDinterwordspacing

\bibitem{Lancers}
\BIBentryALTinterwordspacing
``Lancers co. ltd.'' [Online]. Available: \url{https://www.lancers.jp}
\BIBentrySTDinterwordspacing

\bibitem{schutt2016painfree}
\BIBentryALTinterwordspacing
H.~H. Sch{\"u}tt, S.~Harmeling, J.~H. Macke, and F.~A. Wichmann, ``Painfree and accurate bayesian estimation of psychometric functions for (potentially) overdispersed data,'' \emph{Vision research}, vol. 122, pp. 105--123, 2016. [Online]. Available: \url{https://github.com/wichmann-lab/psignifit/wiki}
\BIBentrySTDinterwordspacing

\bibitem{jorgensen2011predicting}
\BIBentryALTinterwordspacing
S.~J{\o}rgensen and T.~Dau, ``{Predicting speech intelligibility based on the signal-to-noise envelope power ratio after modulation-frequency selective processing.}'' \emph{The Journal of the Acoustical Society of America}, vol. 130, no.~3, pp. 1475--1487, 9 2011. [Online]. Available: \url{http://www.ncbi.nlm.nih.gov/pubmed/21895088}
\BIBentrySTDinterwordspacing

\bibitem{jorgensen2013multi}
\BIBentryALTinterwordspacing
S.~J{\o}rgensen, S.~D. Ewert, and T.~Dau, ``{A multi-resolution envelope-power based model for speech intelligibility},'' \emph{The Journal of the Acoustical Society of America}, vol. 134, no.~1, pp. 436--446, 7 2013. [Online]. Available: \url{http://www.ncbi.nlm.nih.gov/pubmed/23862819}
\BIBentrySTDinterwordspacing

\bibitem{ANSI_S3-5_1969}
A.~S3-5, \emph{{Methods for the calculation of the articulation index}}.\hskip 1em plus 0.5em minus 0.4em\relax Washington DC: American National Standards Institute, 1969.

\bibitem{matsui2022modelling}
T.~Matsui, T.~Irino, R.~Uemura, K.~Yamamoto, H.~Kawahara, and R.~D. Patterson, ``Modelling speaker-size discrimination with voiced and unvoiced speech sounds based on the effect of spectral lift,'' \emph{Speech Communication}, vol. 136, pp. 23--41, 2022.

\bibitem{irino2002segregating}
T.~Irino and R.~D. Patterson, ``Segregating information about the size and shape of the vocal tract using a time-domain auditory model: The stabilised wavelet-mellin transform,'' \emph{Speech Communication}, vol.~36, no. 3-4, pp. 181--203, 2002.

\bibitem{morise2016world}
M.~Morise, F.~Yokomori, and K.~Ozawa, ``World: a vocoder-based high-quality speech synthesis system for real-time applications,'' \emph{IEICE TRANSACTIONS on Information and Systems}, vol.~99, no.~7, pp. 1877--1884, 2016.

\bibitem{github_amlab}
\BIBentryALTinterwordspacing
T.~Irino and K.~Yamamoto, ``Github amlab-wakayama,'' 2021. [Online]. Available: \url{https://github.com/AMLAB-Wakayama/}
\BIBentrySTDinterwordspacing

\end{thebibliography}

%

\end{document}